\documentclass[prl,footinbib,twocolumn,preprintnumbers,amsmath,amssymb]{revtex4-1}

\usepackage{lipsum}
\usepackage{amssymb}
\usepackage{amsmath}
\usepackage{comment}
\usepackage{color}
\usepackage{empheq}
\usepackage{mathtools}
\usepackage{makecell}
\usepackage{enumerate}
\usepackage{mathtools}
\usepackage{stmaryrd}
\usepackage{enumitem}
\usepackage{textcomp}
\usepackage{gensymb}

\newcommand{\moy}[1]{\left\langle #1 \right\rangle}

\newcommand{\dd}[0]{\mathrm{d}}

\newcommand{\rr}[0]{\boldsymbol{r}}

\newcommand{\ff}[0]{\boldsymbol{f}}

\newcommand{\kB}[0]{k_{\mathrm{B}}}

\newcommand{\jj}[0]{\boldsymbol{j}}

\newcommand{\CAB}[0]{E_{A\to B}}
\newcommand{\cAB}[0]{\rho_{A\to B}}
\newcommand{\CBA}[0]{E_{B\to A}}

\newcommand{\rcut}[0]{r_{\mathrm{cut}}}
\newcommand{\rhoAB}[0]{\rho_{A\to B}}
\newcommand{\rhoBA}[0]{\rho_{B\to A}}
\newcommand{\rhoC}[0]{\rho_{C}}
\newcommand{\rhop}[0]{\rho_{p}}
\newcommand{\cC}{\mathcal{C}}
\newcommand{\lbox}[0]{\ell_{\mathrm{box}}}

\definecolor{darkblue}{rgb}{0,0,0.6}
\definecolor{darkred}{rgb}{0.6,0,0}

\begin{document}

\title{Active droplets controlled by enzymatic reactions}

\author{
Jacques Fries$^{1}$, Javier D\'iaz$^{2}$, Marie Jardat$^{1}$, Ignacio Pagonabarraga$^{2}$, Pierre Illien$^{1}$, Vincent Dahirel$^{1}$}

\address{$^{1}$Sorbonne Universit\'e, CNRS, Laboratoire PHENIX (Physico-Chimie des \'Electrolytes et Nanosyst\`emes Interfaciaux (PHENIX)), 4 Place Jussieu, 75005 Paris, France\\
$^{2}$Departament de F\'isica de la Mat\`eria Condensada,
Universitat de Barcelona, Mart\'i i Franqu\'es 1, 08028 Barcelona, Spain and Universitat de Barcelona Institute of Complex Systems (UBICS),
Universitat de Barcelona, 08028 Barcelona, Spain}

\begin{abstract}
The formation of condensates is now considered as a major organization principle of eukaryotic cells. Several studies have recently shown that the properties of these condensates are affected by enzymatic reactions.  
We propose here a simple generic model to study the interplay between two enzyme populations and a two-state protein. In one state, the protein forms condensed droplets through attractive interactions, while in the other state, the proteins remain dispersed. Each enzyme catalyzes the production of one of these two protein states only when reactants are in its vicinity.
A key feature of our model is the explicit representation of enzyme trajectories, capturing the fluctuations in their local concentrations.  The spatially dependent growth rate of droplets naturally arises from the stochastic motion of these explicitly modeled enzymes.
 Using two complementary numerical methods, (1) Brownian Dynamics simulations, and (2) a hybrid method combining Cahn-Hilliard-Cook diffusion equations with Brownian Dynamics for the enzymes, we investigate how enzyme concentration and dynamics influence the evolution with time, and the steady-state number and size of droplets.
Our results show that the concentration and diffusion coefficient of enzymes govern the formation and size-selection of  biocondensates. 
\end{abstract}

\maketitle

\section{Introduction}
Recent developments of imaging techniques below the diffraction limit have shed new light on structural biology at the mesoscale.  A major breakthrough came with the discovery of submicrometer membraneless compartments within cells~\cite{Brangwynne2009,Banani2017}, also called biocondensates. The physical mechanism behind the formation of mesoscale liquid phases in cells has gained much attention in recent years~\cite{Weber2015,Zwicker2022}.

The presence of coexisting droplets in cells suggests that non-equilibrium mechanisms lead to the selection of a specific mesoscale condensate size, arresting Ostwald ripening~\cite{Keber2024}. Continuous descriptions relying on Flory-Huggins free energy have been designed to account for the formation of chemically active droplets~\cite{Zwicker2015, Zwicker2017,Weber2019,Tjhung2018,Ziethen2023}. In these models, active reactions that break detailed balance, and passive ones that respect detailed balance are allowed, with rates that are different inside and outside the droplets~\cite{Zwicker2022, Berthin2024}. These models, solved at a mean-field level, predict stationary states where several droplets coexist. The role of chemical reactions in the formation of biocondensates is consistent with several experimental observations~\cite{Wang2014,Guilhas2020,Linsenmeier2022,Wu2024}. Specifically, post-translational modifications, namely enzyme-catalyzed reactions that change the chemical state of a protein, can modulate the strength of effective interactions between proteins, and therefore either
promote or oppose the formation of biocondensates~\cite{Snead2019}. For instance, two enzymes catalyzing opposite reactions (phosphorylation and dephosphorylation) were correlated to the dynamics of membraneless organelles ({\em e. g.} P Granules in \textit{Caenorhabditis elegans} embryos~\cite{Wang2014}, and postsynaptic condensates in mammalian neurons~\cite{Wu2024}). 

Enzymes catalyzing modifications of condensate proteins may be key controlling agents of the condensate properties~\cite{Perez2024,Smokers2024}. In particular, as condensate are non-equilibrium structures, kinetics of enzymatic reactions and transport properties of enzymes should matter. In this context, several questions emerge. The concentration of each type of enzyme is governed by the genetic metabolism, the kinetics of which leads to a variety of complex dynamical patterns at the heart of systems biology~\cite{Alon2006}. In the field of biocondensates, how does the enzyme concentration qualitatively and quantitatively control the structural properties of the system? Moreover, active mechanisms are known to dramatically affect the diffusion of enzymes towards or within the condensate. What is the influence of the effective diffusion coefficient of enzymes on the phase separation process at play in the formation of biocondensates? 

To address these questions, we propose to model the space and time evolution of biocondensates generated by attractive interactions between proteins, and modulated by 
chemical reactions mediated by enzymes. 
Models describing phase separation and chemical kinetics using Flory-Huggins theory and dynamical equations rely on an artificial density-dependent kinetic rate~\cite{Weber2019,Zwicker2022}. In contrast, we adopt here a more realistic approach. Enzymes are explicitly described by discrete Brownian particles. These enzymes catalyze the switch of a protein from a condensation-prone state, favorable to droplet formation, to a dispersion-prone state.
In our model, these biochemical reactions only occur when enzymes are near the two-state proteins. 
 We consider two types of enzymes, the first one catalyzing a  reaction favoring the dispersion-prone state of the protein, and the second one catalyzing the opposite  reaction leading to the condensation-prone state. 

We resort to two complementary numerical methods to explore the influence of enzyme concentration and dynamics on biocondensates, in 2D :
(1) Reactive all-particle Brownian Dynamics  simulations (these will be called `full BD' in what follows), and (2) A hybrid method (HM) based on Cahn-Hilliard-Cook diffusion equations for the droplet material (the two-state protein), and Brownian Dynamics for the enzymes. Within both simulation schemes, the trajectories of enzymes are explicit. The second method allows us to check the transferability of full Brownian Dynamics  results to more standard mean field models of phase transition dynamics, and to study larger system sizes. 
With both methods, the concentration and the diffusion coefficients of enzymes are shown to influence the number and size of condensates at steady-state. We account for the time evolution of the size of condensates in full BD with a simple analytic model.

\section{Models}

\begin{figure*}
    \centering
    \includegraphics[width=0.9\linewidth]{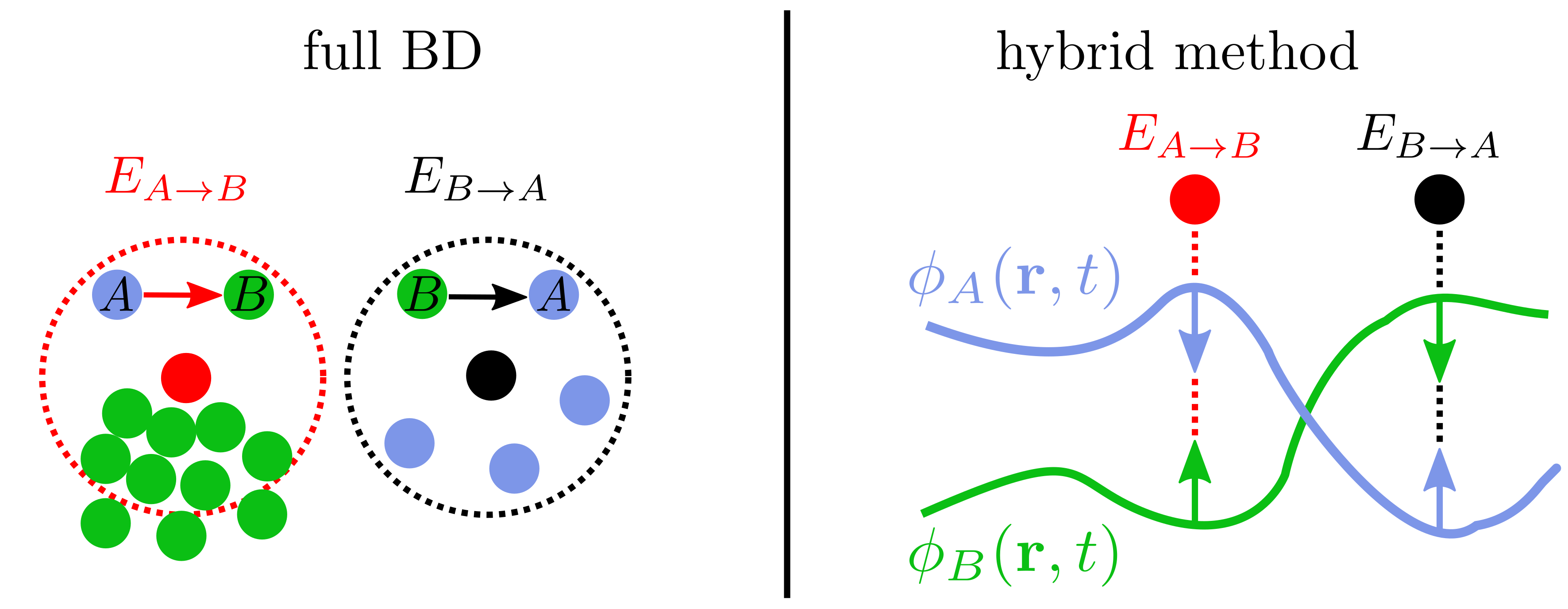}
    \caption{Schematic representation of the principles of the two numerical methods: (left)    full BD, (right) hybrid method (HM) based on Cahn-Hilliard-Cook diffusion equations for the droplet material, and Brownian Dynamics for the enzymes, on the right.}
    \label{fig:BDWCA_hybrid_sketch}
\end{figure*}

We investigate the behavior of a binary mixture of $A$ and $B$ \emph{proteins}, that undergo interconversion reactions $A \leftrightharpoons B$, and where $B$ proteins attract each other and may form droplets, representing the biocondensates. The reactions $A\to B$ and $B\to A$ are respectively catalyzed  by \emph{enzymes} called $\CAB$ and $\CBA$. We assume that these reactions do not take place without enzymes. These enzymes are explicitly represented, but with a highly coarse-grained representation, as disks. Note that these two reactions are coarse-grained representations of processes that involve hidden chemical reactions. Indeed, in a biological context, one of these two reactions would be coupled to a favorable secondary reaction, such as ATP hydrolysis. 
Moreover, as for all catalysts, an enzyme should facilitate the reactions in both ways (in our case, $A\to B$ and $B\to A$). 
For each one of the two reactions catalyzed by $\CAB$ and $\CBA$ enzymes, we consider that the reverse reaction is much less likely than the forward reaction, 
and is thus neglected (this assumption is justified in appendix \ref{AppendixBD}). 
We restrict ourselves to two-dimensional systems with periodic boundary conditions.  

In full Brownian Dynamics, $A$ and $B$ proteins are explicitely represented. The number of enzymes is varied from one simulation to another, and these particles are replaced if necessary by non-catalyzing {\em neutral} particles $C$ to keep the total number of particles $N$, and thus the overall surface fraction, constant. We denote by $S_i(t) \in \{A, B, C, \CAB, \CBA\}$ the species of particle $i$ at time $t$.  The positions of particles satisfy the overdamped~\cite{Ermak1975}:
\begin{equation}
\label{DBsuramortie }
\frac{\dd\rr_i}{\dd t} = - \frac{D_i}{\kB T} \sum_{j \neq i }\nabla U_{S_i,S_j}(|\rr_i-\rr_j|) + \sqrt{2 D_i} \boldsymbol{\eta}_i(t),
\end{equation}
where 
$D_i$ is the bare diffusion coefficient of particle $i$, and $\boldsymbol{\eta}_i(t)$ is a white noise such that $\moy{\eta_{i,\alpha}(t)}=0$ and $\moy{\eta_{i,\alpha}(t)\eta_{j,\beta}(t')}=2D_i \delta_{ij} \delta_{\alpha\beta} \delta (t-t')$ for any components $\alpha,\beta = x$ or $y$. All particles interact through a repulsive Weeks-Chandler-Andersen (WCA) potential~\cite{Weeks1971}, except $B$ proteins that interact through a Lennard-Jones (LJ) potential. All particles have the same diameter $\sigma$, which sets the length scale. Similarly, the surface densities of all the species are measured in units of $\sigma^{-2}$ All particles have the same diffusion coefficient $D_i=D_0$.
The WCA potential reads 
\begin{equation}
    {U_{S_i,S_j}
    (r_{ij}) }= 
    \begin{cases}
     4\varepsilon' \left[ \left(\frac{\sigma}{r_{ij}}\right)^{12}-\left(\frac{\sigma}{r_{ij}}\right)^{6}  \right] + \varepsilon' & \text{if $r_{ij} \leq 2^{1/6}\sigma$,} \\
     0& \text{otherwise,}
    \end{cases}
\end{equation}
for any couple $S_i,S_j$ except $S_i=S_j=B$, where $r_{ij}$ is the distance between $i$ and $j$. We take $\varepsilon'=10k_{\rm B}T$. 
The Lennard-Jones potential between $B$ proteins includes an attractive part:
\begin{equation}
   {U_{BB}
   (r_{ij}) }= 4\varepsilon \left[ \left(\frac{\sigma}{r_{ij}}\right)^{12}-\left(\frac{\sigma}{r_{ij}}\right)^{6}  \right] ,
\end{equation}
 for which we set a cutoff for $r_{ij} \geq 2.5 \sigma$.
The depth of the LJ potential energy well is $\varepsilon = 3k_{\rm B}T$, \emph{i.e.} below the critical temperature for vapor-liquid phase equilibrium of the two-dimensional Lennard-Jones fluid~\cite{Smit1991}, enabling $B$ proteins to form phases of contrasting densities.

Reactions are introduced in the algorithm through a random telegraph model~\cite{van1976stochastic}, parameterized by a reaction time $1/k$. Theses reactions are only allowed in the vicinity of the enzymes~\cite{Decayeux2021,Decayeux2023}: when a protein of type $A$ (resp. $B$) is at a distance smaller than a cutoff distance $r_\text{cut}$ from the center of the  $\CAB$ enzyme (resp. $\CBA$), it becomes $B$ (resp. $A$) with rate $k$ (Fig. \ref{fig:BDWCA_hybrid_sketch}). Therefore, the local reaction rate is coupled to the random trajectory of the Brownian enzymes. The relationship of our model with typical biological situations is specified in appendix \ref{AppendixBD}, which specifies why our model is equivalent to a mixture of two enzymes respectively catalyzing a passive and an active reaction.  

In the hybrid method, the positions of enzymes evolve through the overdamped Langevin equation of Brownian Dynamics, and are coupled to a binary $A$-$B$ fluid evolving through a continuous diffusion equation~\cite{Cahn1959}. The fluid is characterised by the order parameter $\psi(\rr,t)=\phi_A-\phi_B$ defined as the difference in surface fraction of species $A$ and $B$ ($A$ and $B$ cover all space , so $\phi_A + \phi_B = 1$ and thus $-1 \le \psi \le 1$). The free energy of the $A$-$B$ fluid is given by a standard Ginzburg-Landau density functional~\cite{Cahn1959}. The parameters are chosen so as to favor phase separation. The dynamics of the $A$-$B$ fluid is controlled by the Cahn-Hilliard-Cook standard equation, with the addition of reactive fluxes that take place in the vicinity of each Brownian enzyme (Fig.~\ref{fig:BDWCA_hybrid_sketch}). 

Details on the methods,  and values of the parameters are given in appendix \ref{AppendixBD} for full BD, and appendix \ref{appendixHM} for the hybrid method.

\section{Arrested phase separation and emergence of non-equilibrium structures}

\begin{figure*}
    \centering

\includegraphics[width=0.8\textwidth]{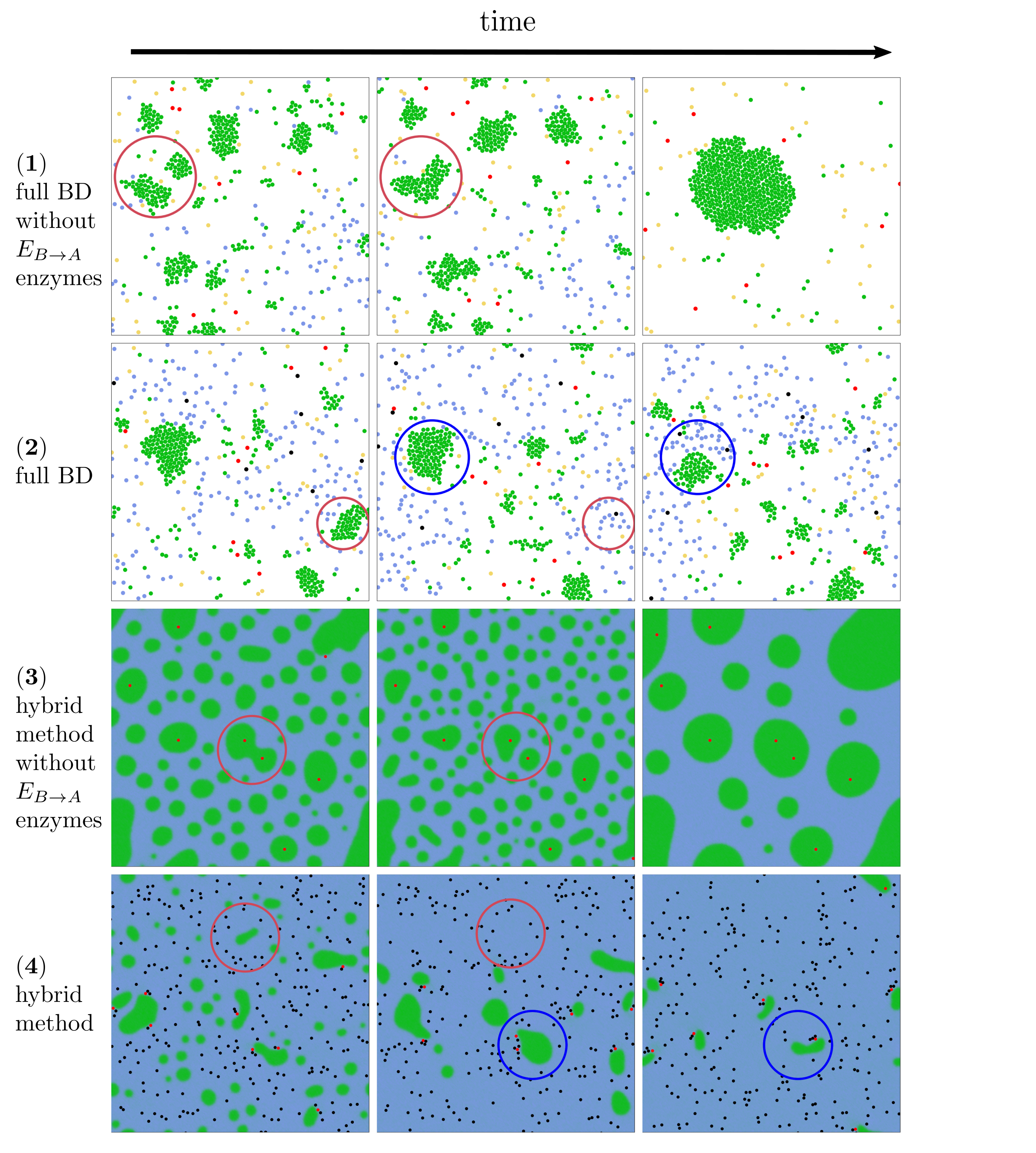}
    \caption{\textbf{Droplet growth under the effect of active enzymatic reactions.} In all cases, discrete particles are represented by small disks: $\CAB$ enzymes are colored in red, $\CBA$ enzymes are colored in black. \textbf{(1 and 2)} Snapshots from trajectories obtained with full BD, at successive times. $A$ proteins are colored in blue, $B$ proteins  are colored in green, neutral particles $C$ are colored in orange with the surface concentrations of enzymes equal to $\rhoAB=1.6\ 10^{-3}$ (and $\rhoBA=1.6\ 10^{-3}$ in $\bf 2$). \textbf{(3 and 4)} Snapshots obtained with the hybrid method. The continuous phase is colored in blue as it plays the role of the $A$-rich phase, while the discontinuous $B$-rich phase is colored in green. Here with $\rhoAB=8.8\ 10^{-5}$ (and $\rhoBA=2.8\ 10^{-3}$ in $\bf 4$).  
    \textbf{(1 and 3)}: Reference system with $\CAB$ enzymes only that leads to the uninterrupted growth of droplets (note that the trajectory of the HM method is not at stationary state yet);  \textbf{(2 and 4)}: System containing $\CAB$ and $\CBA$ enzymes that leads to the interrupted growth of droplets.
    Red and blue circles are drawn to attract attention to certain noteworthy events, which are detailed in the body of the article.}
    \label{fig_2_pictures}
\end{figure*}


As a reference, we consider a system containing only $\CAB$ enzymes, starting with an initial situation with only $A$ proteins in the system. 
 In this situation, a phase separation occurs with the uninterrupted growth of droplets of $B$ proteins, as seen both in full BD and in hybrid  simulations (Fig.~\ref{fig_2_pictures}-\textbf{1} and \ref{fig_2_pictures}-\textbf{3}).  First, small droplets of $B$ proteins are formed near the $\CAB$ enzymes, and second, the largest of these droplets of $B$ material grow at the expense of the smallest, by Ostwald ripening or by coalescence. Coalescence events are depicted in the red circles of Figs.~\ref{fig_2_pictures}-\textbf{1} and \ref{fig_2_pictures}-\textbf{3}.
 In full BD simulations, a single droplet of $B$ protein is observed at stationary state. One difference between the results of the two simulation methods is that in full BD, the $\CAB$
enzymes continue to diffuse freely within the simulation box, even after some droplets have nucleated, whereas in the hybrid method, they remain attached to the 
$B$ droplets.

In the presence of both types of $\CAB$ and $\CBA$ enzymes, droplet growth is interrupted, as observed in both simulation schemes. Indeed, $\CBA$ enzymes, by converting $B$ proteins into $A$ ones, either limit the growth (as depicted in the red circles of Figs.~\ref{fig_2_pictures}-\textbf{2} and \ref{fig_2_pictures}-\textbf{4}), or completely destroy the droplets (as depicted in the blue circles of the same figures). As we proceed to show, the balancing effect of the two types of enzymes results in the selection of a droplet size. 

\section{The concentration of enzymes controls the size and number of droplets}

\begin{figure*}
    \centering 
  \includegraphics[width=1\linewidth]{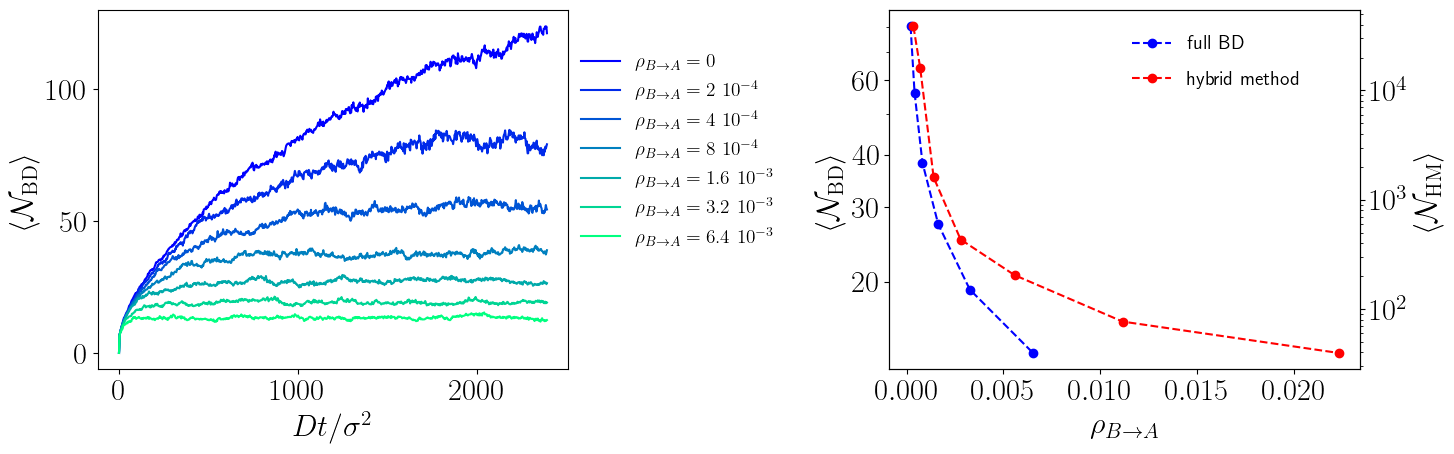}
\caption{\textbf{Enzymes arrest droplet growth}. Left: Average number of $B$ proteins in a droplet obtained by full BD as a function of time for increasing values of the surface concentration $\rhoBA$ of $\CBA$ enzymes.  
Right: Average size of droplets at stationary state obtained by full BD and by the hybrid method as a function of the surface concentration of $\CBA$ enzymes, in log-scale.
In full BD,  the concentration of $\CAB$  enzymes is $\cAB = 1.6~10^{-3}$, and the stationary state is assumed to be reached in the interval $[1800,2400]\frac{Dt}{\sigma^2}$. In the HM,
the surface concentration of $\CAB$  enzymes is $\cAB = 1.2~10^{-4}$,  and the stationary state is assumed to be reached in the interval $[5~10^4,12.5~10^4]t$.}
\label{fig_3_dp_growth}
\end{figure*}

\begin{figure*}
    \centering 
  \includegraphics[width=1\linewidth]{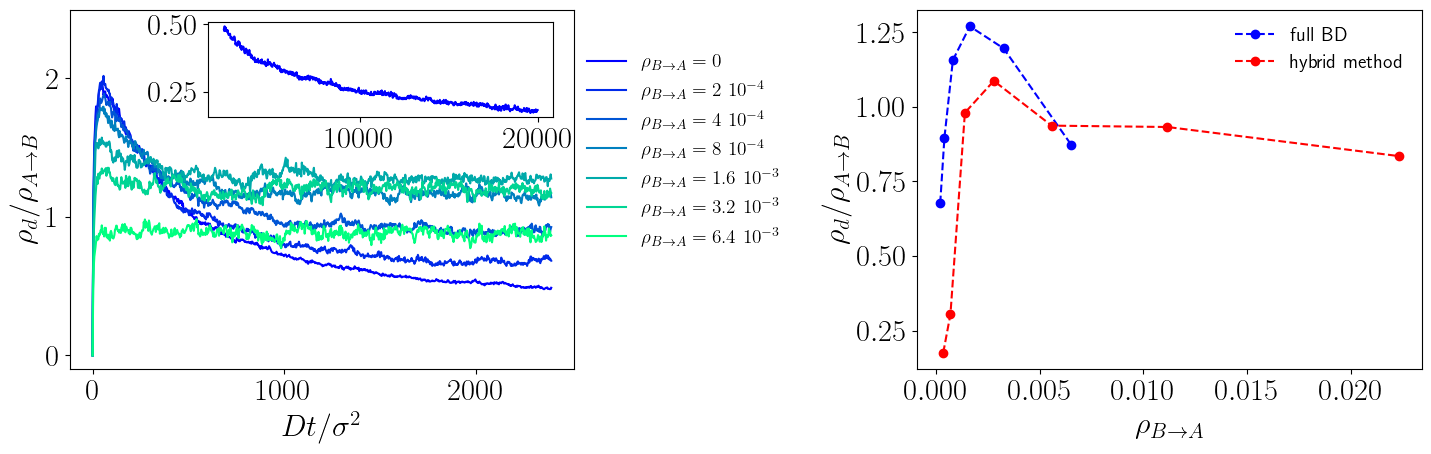}
\caption{\textbf{Non-monotonic influence of the enzyme concentration on the number of droplets.}
Left: Average number of droplets per $\CAB$ enzyme, measured by the ratio of the concentration of droplets $\rho_d$ to the concentration $\rhoAB$ of $\CAB$ enzymes, as a function of time, obtained by full BD for increasing values of the surface concentration $\rhoBA$ of $\CBA$ enzymes. The inset is the prolongation of the curve with $\rhoBA=0$ on the interval $[2400,20000]\frac{Dt}{\sigma^2}$. The y scale is the same as in the main plot. Right: Average number of droplet per enzyme $\CAB$ at stationary state obtained by {\em full BD} and by the hybrid method as functions of the surface concentration $\rhoBA$ of $\CBA$ enzymes. In full BD, the concentration of $\CAB$  enzymes is $\cAB = 1.6~10^{-3}$, and the stationary state is assumed to be reached in the interval $[1800,2400]\frac{Dt}{\sigma^2}$. In the HM,
 the surface concentration of $\CAB$  enzymes is $\cAB = 1.2~10^{-4}$,  and the stationary state is assumed to be reached in the interval $[5~10^4,12.5~10^4]t$.}

\label{fig_4_average_nbr_droplet}
\end{figure*}

As shown in Fig.~\ref{fig_3_dp_growth}-left from full BD simulations, at a fixed surface concentration $\rhoAB$ of $\CAB$ enzymes, a non-zero concentration of $\CBA$ enzymes leads to the formation of droplets that reach a finite size at stationary state.  Moreover, the size of these droplets, measured by the number of $B$ proteins in a droplet $\mathcal{N}_{\mathrm{BD}}$, decreases as the surface concentration of $\CBA$ enzymes increases. As previously stated, in the absence of $\CBA$ enzymes ($\rhoBA=0$, blue plot), we observe the growth of a unique droplet of $B$ proteins (though the equilibrium state is not reached here, as the simulation is too short). The same general behaviour is captured by the hybrid method, as shown in Fig.~\ref{fig_3_dp_growth}-right, where we plot the evolution of the average size of $B$ droplets as a function of the surface concentration of $\CBA$ enzymes, obtained from both numerical methods, on a logarithmic scale. The surface concentration of $\CAB$ is $10$ times smaller in the HM than in full BD, leading to quantitative differences in the results. Note also that in the HM, a droplet is a cluster of contiguous cells where the concentration of $B$ dominates, and $\mathcal{N}_{\mathrm{HM}}$ is proportional to the number of contiguous cells, as detailed in Appendix \ref{appendixHM}. In contrast, a droplet in full BD is a cluster of more than $5$ $B$ proteins, as detailed in Appendix \ref{AppendixBD}.

As the presence of $\CBA$ enzymes interrupts Ostwald ripening, multiple droplets may coexist at stationary state, whose number depends on the surface concentration $\rhoBA$ of $\CBA$ enzymes, as shown in Fig.~\ref{fig_4_average_nbr_droplet}-left from full BD simulations at a fixed surface concentration $\rhoAB$ of $\CAB$ enzymes. In the absence of $\CBA$ enzymes ($\rhoBA=0$, blue plot), we expect a ratio $\rho_d/\rhoAB$ of $0.125$ at long times, as the number of $\CAB$ enzymes is $8$ in the simulation box for this $\cAB$ value, and a single droplet is expected at equilibrium (although the simulation time here is too short to reach this state). In all other cases, a stationary state is reached, with a mean number of droplets per $\CAB$ enzyme larger than $0.125$, indicating the formation of multiple droplets. In some cases, the number of droplets per $\CAB$ enzyme exceeds $1$. Interestingly, both numerical methods show a non-monotonic evolution of the number of droplets per $\CAB$ enzyme, with a maximum appearing at roughly the same $\CBA$ enzyme concentration (but for different values of $\rhoAB$), as shown in Fig.~\ref{fig_4_average_nbr_droplet}-right. 
This non-monotonic behavior arises because, at low concentrations of $\CBA$ enzymes, the number of droplets is expected to increase with $\rhoBA$ from an initial value equal to $1$ at $\rhoBA=0$ (equilibrium state). Indeed, as soon as some $\CBA$ enzymes are introduced, the production of new $A$ proteins enables the nucleation of new droplets in the vicinity of $\CAB$ enzymes. However, in the limit of high $\CBA$ enzyme concentration, we expect the immediate destruction of droplets after nucleation, resulting in the disappearance of all droplets. As a result, the number of droplets passes through a maximum for an intermediate value of $\rhoBA$.
There are quantitative differences between both simulation results at large $\rhoBA$ values, where the decrease of the number of droplets is less pronounced with the hybrid method. 
With this method, B-rich droplets are always physically connected to a least one $\CAB$ enzyme. This connection stabilizes droplets down to small sizes, even if the droplet encounters an $\CBA$ enzyme. In full BD simulations, $\CAB$ enzymes may diffuse away from droplets, thus making these condensates more likely to be fully destroyed under the influence of the reactions catalyzed by an $\CBA$ enzyme.

\begin{figure*}
  \includegraphics[width=1\linewidth]{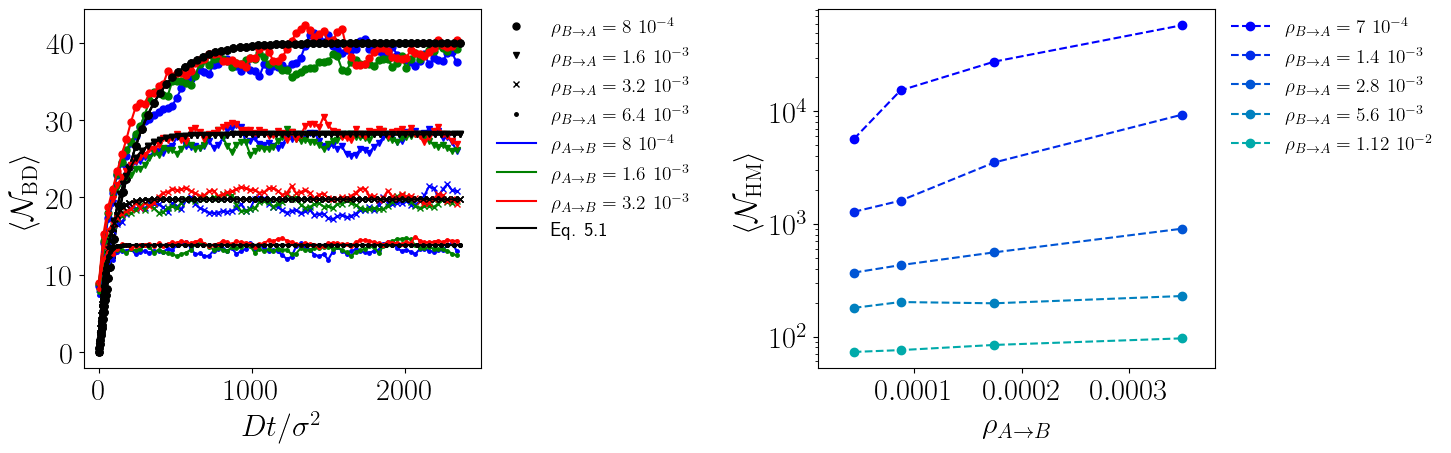}
  \caption{\textbf{Influence of the concentration of $\CAB$ enzymes.}
  Left: Average number of $B$ proteins in a droplet obtained by full BD as a function of time for several values of the surface concentration $\rhoBA$ of $\CBA$ enzymes, and several values of the surface concentration $\rhoAB$ of $\CAB$ enzymes.  The results obtained by Eq.~\ref{expression_analytique_N} are displayed in plain black with parameters $\alpha=2.7$ and $\beta=900$.  
Right: Average size of droplets at stationary state obtained by  the hybrid method as a function of the surface concentration of $\CAB$ enzymes for increasing values of the concentration $\rhoBA$ of $\CBA$ enzymes, in log-scale.
}
  \label{fig_5_concentration_Eab}
\end{figure*}

In the range of parameters explored by full BD, it appears that the concentration of $\CAB$ enzymes does not affect the droplet size (see Fig.~\ref{fig_5_concentration_Eab}-left), while an increase in the concentration of $\CBA$ enzymes leads to a decrease in droplet size, as previously mentioned. This suggests that the droplets grow independently from each other around each $\CAB$ enzyme, until they encounter an $\CBA$ enzyme. The results from the HM, displayed in Fig.~\ref{fig_5_concentration_Eab}-right, qualitatively follow the same trend, but quantitatively differ, with the size of droplets increasing more rapidly as a function of the concentration of $\CAB$ enzymes (note the logarithmic scale of this plot). This is likely due to the fact that the regime of $\CAB$ enzyme concentration explored using the hybrid method differs from that examined in the full BD simulation (the concentrations of $\CAB$ are significantly lower with the HM). This is made possible because the HM allows for the study of much larger systems than the full BD. 
All in all, our results show that the size of the droplets are tuned by enzyme concentrations.

\section{Enzyme diffusivity affects the size of condensates}

When they are chemically active, the effective diffusion coefficient of enzymes can actually strongly differ from their equilibrium value, typically given by the Stokes-Einstein relation. This idea originates from the numerous sets of measurements in the biophysical literature, together with  theoretical explanations which aim at relating the chemical activity of the enzyme with their diffusivity~\cite{Agudo2018, Feng2020, Ghosh2021}.

\begin{figure*}
\includegraphics[width=1 \linewidth]{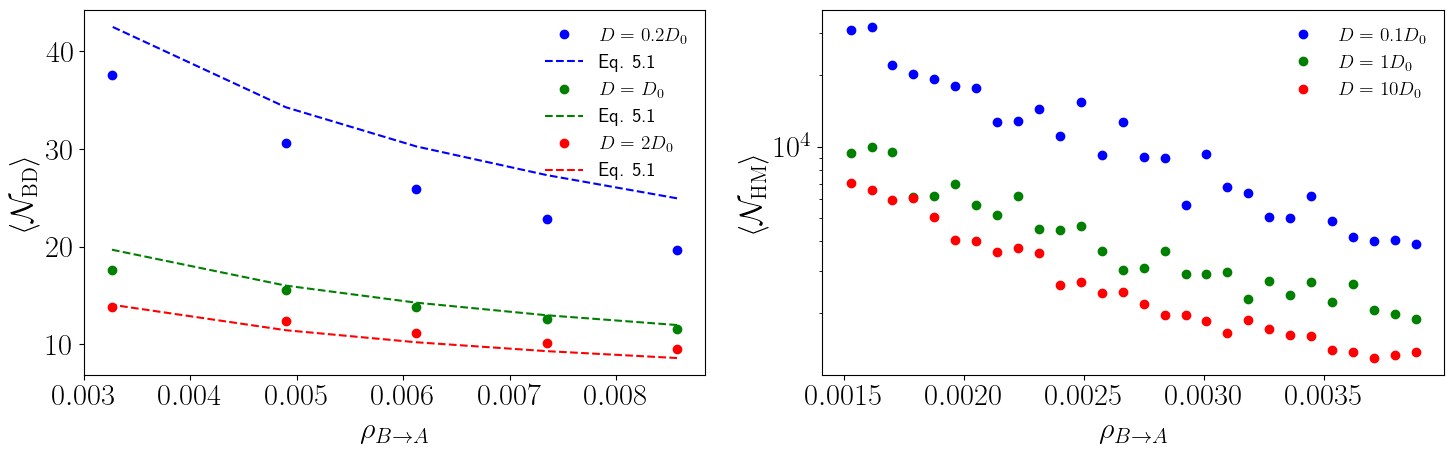}
\caption{\textbf{Influence of the diffusion coefficient of enzymes on the droplet size. } Left: Average number of $B$ proteins in a droplet at stationary state obtained by full BD for several values of the diffusion coefficient of $\CBA$ and $\CAB$ enzymes, as functions of the surface concentration $\rhoBA$ of $\CBA$ enzymes.  The results obtained by Eq.~\ref{expression_analytique_N} are displayed in dashed lines with parameters $\alpha=2.7$ and $\beta=900$. Here $\rhoAB=1.6\ 10^{-3}$.  
Right: Average size of droplets at stationary state obtained by the hybrid method as a function of the surface concentration of $\CAB$ enzymes for several values of the diffusion coefficient of enzymes $\CBA$ and $\CAB$, as functions of the surface concentration $\rhoBA$ of $\CBA$ enzymes, in log-scale. Here, $\rhoAB=2.4\ 10^{-4}$.}
\label{fig_6_vitesse_enzyme}
\end{figure*}

The diffusion coefficient of particles is a parameter of Brownian Dynamics simulations, and can be easily modified. Here, this value for $\CAB$ and $\CBA$ enzymes is either enhanced or decreased compared to the value $D_0$ previously used, keeping the same size for all BD particles, and the dynamic properties of all other species unchanged.
As it appears in Fig.~\ref{fig_6_vitesse_enzyme}, the average droplet size is strongly influenced by the diffusion coefficient of enzymes: the faster are the enzymes, and the smaller are the droplets. This behavior is observed in both simulation methods. 
This trend may be related to the encounter time between enzymes and droplets: decreasing this time increases the effective rate of interconversion of $A$/$B$ species. 
 
We propose in Appendix~\ref{annex_analytical_model} a simple analytic model that accounts for this behaviour. The assumptions of this model are: (i) each time a droplet encounters an $\CBA$ enzyme, it is instantaneously destroyed, (ii) the encounter times between droplets and $\CBA$ enzymes follow an exponential probability distribution, (iii) the encounter time $t_e$ between a droplet and a single enzyme only depends on the diffusion coefficients of both entities. With these assumptions, we obtain the evolution with time of the average number of particles per droplet $ \langle \mathcal{N} \rangle$:
\begin{align}
    \langle \mathcal{N} \rangle
    =c \frac{t_e}{\mathcal{S}\rho_{B\to A}} \left(1-\exp\left(\frac{-t \mathcal{S} \rho_{B\to A}}{t_e}\right)\right),
    \label{expression_analytique_N}
\end{align}
with $\mathcal{S}$ the surface of the simulation box, $c$ a constant that represents the influx of protein into a droplet. The encounter time $t_e$ is assumed to be inversely proportional to the sum of the diffusion coefficients of the droplet $D_{\rm dp}$ and of the $\CBA$ enzyme $D_{\CBA}$:
\begin{align}
    t_e = \frac{\beta}{(D_{\CBA}+D_{\mathrm{dp}})}.
    \label{eq_time_encounter}
\end{align}
The full BD simulation data shown in Figs. \ref{fig_5_concentration_Eab}-left and \ref{fig_6_vitesse_enzyme}-left are successfully fitted by Eq.~\ref{expression_analytique_N}, provided that we take
\begin{align}
    c = \alpha \sqrt{\frac{\mathcal{S}\rhoBA}{t_e}},
    \label{eq_time_encounter}
\end{align}
\begin{align}
    D_{\mathrm{dp}}=\frac{D_0}{{\langle \mathcal{N} \rangle_{t\to\infty}}}.
    \label{eq_time_encounter}
\end{align}
The dependence of $c$ with $\mathcal{S}\rhoBA$ is consistent with the fact that 
the presence of $\CBA$ enzymes increases the amount of $A$ proteins in the system. A higher concentration of $A$ increases the influx of proteins in the droplet.
The dependence of $D_{\mathrm{dp}}$ with the size of droplets at stationary state is justified by the fact that the diffusion coefficient of the center of mass of an assembly of $N$ Brownian particles is inversely proportional to the number of particles. 
The  agreement of this simple analytic model with full BD simulation results proves the major role played by the encounter time between a droplet and an $\CBA$ enzyme to control the size of droplets. In the case $D = 0.2 D_0$, the comparison of simulation results with the model would work better with a slightly higher diffusion coefficient of the enzymes. 
Interestingly, this is consistent with calculations of the mean square displacements of the enzymes in these cases, which show a slight increase of the diffusion coefficient. 
We think that the interface between the two phases where the enzymes are located may push the enzymes and lead to self-propulsion. In previous studies, we have already characterize similar behaviors in a model of colloidal particle propelled by a finite size domain of Lennard-Jones particles~\cite{Decayeux2021,Decayeux2023}. 
Despite this unusual behavior, that we have to characterize in more details in future works, the ability of the model to fill all the full BD simulation data is remarkable.

The HM also shows that larger enzyme diffusivity leads to smaller droplet sizes. Despite the qualitative agreement, the analytical model does not quantitative account for the results obtained in the regime of large systems, described with the HM. From the dynamics point of view, the hybrid simulations show two singularities that explain this discrepancy with the analytical model. First, the mass transport of the $A$/$B$ fluid does not limit the rate at which the droplet grows, as the droplet interface follows the motion of the $\CAB$ enzymes: in all snapshots, we observe that the $\CAB$ particles remain in contact with this interface.
Secondly, a first investigation of the mean squared displacements of the enzymes suggests a strong self-propulsion in this range of parameters, which may dominate normal diffusion.  
Lastly, in the HM model, some enzymes are found inside droplets, which reflects the enzyme composition of real biocondensates, but that we do not account for in the simple analytical model.

\section{Conclusion}

In this work, we developed a simple generic model to study the interplay between two enzyme populations and a two-state protein. It is well known that biological systems regulate properties at the subcellular scale by controlling the spatial and temporal distribution of enzyme density. Enzymes selectively determine how biological components are modified and can thereby alter the physical interactions between their substrate proteins. A significant portion of enzyme-catalyzed reactions are coupled with highly exergonic processes, such as ATP hydrolysis, allowing active reactions to drive product molecules or mesoscale condensates far from equilibrium.

Here, we model a two-state protein, assumed to be the substrate of two distinct enzymes. In one state, the protein (state $B$) forms condensates through attractive interactions, while in the other state (state $A$), the proteins remain dispersed. Each enzyme catalyzes the production of one of these two protein states, promoting either protein clustering ($\CAB$ enzyme) or dispersion ($\CBA$ enzyme). Using this simple model, and employing two different simulation methods at different scales, we have demonstrated that enzyme concentration and diffusion coefficients govern the size and number of biocondensates (droplets of $B$ protein).
A key feature of our model is the explicit representation of enzyme trajectories, capturing the fluctuations in their local concentrations. Unlike other models of biocondensates that rely on density-dependent reaction rates to induce size-selected droplets~\cite{Zwicker2022}, our approach does not require such assumptions. The spatially dependent growth rate of droplets naturally arises from the stochastic motion of explicitly modeled enzymes.

Our minimal model suggests key design principles for enzymatic systems that regulate biocondensate properties. It opens the way to further studies integrating concepts from systems biology (networks of enzymatic reactions), macromolecule transport phenomena in biological fluids, to the dynamics of liquid-liquid phase separation at mesoscales.

\appendix

\section{APPENDICES}

\section{A1. Methods: All-particle Brownian Dynamics simulations (full BD)}
\label{AppendixBD}
\subsection{A1a. Simulation parameters}

We consider two-dimensional systems with periodic boundary conditions, that contain $A$ and $B$ proteins, $\CAB$ and $\CBA$ enzymes and $C_N$ neutral particles, also referred to as crowders. 
The total number of particles is fixed, ensuring the same overall density. Specifically, the number of $A$ and $B$ proteins is fixed, while the number of  $\CAB$ and $\CBA$ enzymes varies. The total particle number is adjusted using neutral particles $C_N$.
All particles have the same diameter $\sigma$, which is used as the unit length. $D_0$ denotes the bare diffusion coefficient of the particles, and $\sigma^2/D_0$ is used as a unit time.
In all cases, the simulation box, with a length $\lbox =70 \sigma$ contains a surface concentration of enzymes and crowders $\rhoAB+\rhoBA+\rhoC=0.01$, and a concentration of $A$ and $B$ proteins, $\rho_S=0.1$. This corresponds to $50$ enzymes and crowders, and $500$ $A$/$B$ proteins. The integration time step of the overdamped Langevin equation is $\Delta t=1\cdot 10^{-4}\sigma^2/D_0$ in all cases.

The conversions of $A$ and $B$ proteins take place whenever they are within a distance $\rcut$ to the center of an enzyme ($\rcut=5$ is taken). 
 More precisely, at each time step, in the vicinity of an enzyme $\CAB$ (resp. $\CBA$), an $A$ protein may transform into $B$ at rate $k_{A\to B}$ (resp. a $B$ protein may transform into $A$ at rate $k_{B\to A}$). All the other reverse reactions are neglected.
In all simulations, we assume that the reactions are fast compared to the diffusion characteristic time scale, and we take $k_{A\to B}  = k_{B\to A} =10$. The choice of the integration time step ensures that $k_{A\to B} \Delta t $ and $k_{B\to A} \Delta t $ remain much smaller than $1$.

The characterization of the size and of the number of droplets is done using a Voronoi cell analysis. 
We found that the distribution of the size of Voronoi cells around $B$ proteins is bimodal, which allows us to define a threshold below which a particle can be tagged as being part of a droplet.    The square root of this threshold, $2\sigma$, is taken as the distance criteria to identify particles belonging to the same droplet. A droplet is then considered as a group composed of more than $5$ $B$ particles. From such analysis, the distribution of the number of droplets, and of the number of particles per droplet is computed, as well as moments of these distributions over time, and at stationary state. In each case, the results  are averaged over $50$ independent realisations of the Brownian trajectories.

\subsection{A1b. Interpreting our model : passive and active reactions in living systems}

In this section, we aim at clarifying how the two kinds of chemical reactions described by our model can be considered as a couple of passive and active reactions. 
The term \emph{active} has been used in the context of cellular metabolism for more than half a century~\cite{Szent1958,Jencks1989}.
From a qualitative point of view, in biological systems, some reactions, referred to as \emph{active reactions},
maintain the system far from equilibrium: under the influence of an active reaction $A \rightleftharpoons B$, 
the evolution of the system composition does not relax towards chemical equilibrium ($\mu_A = \mu_B$). Instead, at a specific enzymatic site, the $A \rightleftharpoons B$ reaction is coupled to another process associated with a negative free energy (that we shall call \emph{chemical drive}~\cite{Zwicker2015}, noted $\Delta \mu$). This energy source may be another chemical reaction (such as ATP hydrolysis), but also the transport of particles associated with electrostatic or osmotic works~\cite{Mitchell1974}. 
As a consequence of this coupling, the evolution of the system under the effect of the active reaction $A \rightleftharpoons B$ drives the system to a non-equilibrium state characterized by $\mu_B - \mu_A = \Delta \mu$. 

In biological systems, the same particles may be implicated in both passive and active reactions. Nevertheless, these two reactions are catalyzed by distinct enzymes. For instance, a phosphatase (a \emph{passive} enzyme) catalyses the \emph{passive} hydrolysis of a phosphate group bound to a protein residue, while a phosphorylase (an \emph{active} enzyme) catalyzes the reverse \emph{active} phosphorylation of the same protein residue, coupling this protein modification to ATP hydrolysis~\cite{Shacter1984,Snead2019}. 

From a microscopic point of view, activity breaks the detailed balance rules that constrain the transition probabilities at play in these reactions~\cite{Zwicker2015,Berthin2024}. To clarify this point in the context of Brownian Dynamics simulations, we consider two configurations $\mathcal{C}$ and $\mathcal{C}'$ of respective energies $E(\cC)$ and $E(\cC')$, at two successive time steps (\emph{i.e.} at times $t$ and $t+\delta t$). These configurations only differ by the state of one particle : $\mathcal{C}$ changes to $\mathcal{C}'$ with a probability $k_{A\to B} \delta t$ as a result of one iteration of the forward $A \to B$ reaction, 
while $\mathcal{C}'$ changes to $\mathcal{C}$ with a probability $k_{B\to A} \delta t$. 

When the system is at equilibrium, interconversions take place through a \emph{passive} pathway. 
The \emph{passive} conversion rates $k^{\text{p}}_{A\to B}$ and $k^{\text{p}}_{B\to A}$ obey the detailed balance condition:   

\begin{equation}
\frac{k^{\text{p}}_{A\to B}}{k^{\text{p}}_{B\to A}} = \exp\{-\beta[E(\cC')-E(\cC)]\}
\end{equation}

where $\beta=(\kB T)^{-1}$. In non-equilibrium systems, alternative active pathways may exist. In this case, 
the \emph{active} conversion rates $k^{\text{a}}_{B\to A}$ and $k^{\text{a}}_{A\to B}$ break the detailed balance condition:   
\begin{equation}
\frac{k^{\text{a}}_{A\to B}}{k^{\text{a}}_{B\to A}} = \exp\{-\beta[E(\cC')-E(\cC)+\Delta \mu]\}
\end{equation}

The chemical drive $\Delta \mu$ quantifies the deviation from equilibrium. 
Starting from these rules, we then explicit the specific regime corresponding to our model, with $\CAB$ enzymes catalyzing only the forward reaction $A \to B$, and $\CBA$ enzymes catalyzing only the reverse reaction. 

The energies $E(\cC)$ and $E(\cC')$ are a combination of two terms, (1) the sum of interparticle interactions, which depends on the particle coordinates and (2) the intraparticle (or internal) free energy. 
We work in a regime where the difference in the internal free energies of the $A$ and $B$ species are much larger than $\kB T$. This quantity can be identified with the standard reaction free energy for $A \rightleftharpoons B$, which has been estimated for many biological reactions and whose absolute value usually lies in the range $10-20$ kJ.mol$^{-1}$, \emph{i.e.} $5$ to $10$ $\kB T$~\cite{Alberty2000}. This leads to  several simplifications.

First, when a reaction occurs, the change in internal free energies is much larger than the 
contribution of interparticle interactions, which can thus be ignored. 
The energy difference then reads $E(\cC')-E(\cC) = w_B - w_A = \Delta w$, where $w_A$ and $w_B$  are the respective internal free energies of particles $A$ and $B$. 

If the passive reaction is favored in the $A \to B$ direction, then $\Delta w \ll -k_B T$, $\exp\{-\beta[E(\cC')-E(\cC)]\} \gg 1$, and $k^{\text{p}}_{B\to A} \ll k^{\text{p}}_{A\to B}$. As we proceed to show in the next Appendix section, in such case the $B \to A$ reactions can be neglected. In our model, this would then correspond to the rules governing the transitions of $A$ and $B$ particles in a region close to the $\CAB$ enzyme, which could then be considered as a \emph{passive} region. 

Moreover, in biological systems, the active and passive pathways drive the system composition in opposite directions. In other words, the chemical drive is large enough to reverse the direction of the passive reactive flux. In the case of ATP hydrolysis as a source of chemical drive, $\Delta \mu = \mu_{\rm ATP} - \mu_{\rm ADP}$. This quantity is in the range $-40-60$ kJ.mol$^{-1}$, \emph{i.e.} about $-25 \kB T$. 
In the case $\Delta w \ll -k_B T$, the later conditions leads to $\Delta \mu \ll \Delta w \ll -k_B T$. 
This implies that $k^{\text{a}}_{B\to A} \gg k^{\text{p}}_{A\to B}$. The $A \to B$ reactions can be neglected. In our model, this would then correspond to the region close to the $\CBA$ enzyme, which could then be considered as an \emph{active} region. 

The symmetrical case, $\Delta w \ll k_B T$ leads to $k^{\text{p}}_{B\to A} \ll k^{\text{p}}_{A\to B}$. A reverse active reaction flux occurs for $\Delta \mu \ll \Delta w \ll k_B T$.
This implies that $k^{\text{a}}_{A\to B} \ll k^{\text{a}}_{B\to A}$. Under this scenario, $\CAB$ would catalyze the active pathway, and $\CBA$ the passive one. 

All in all, our model is consistent with the coexistence of active and passive regions around two kinds of Brownian enzymes that drive the chemical composition of the system in opposite directions. The assumptions for the values of the chemical drive and for the internal free energies of the reactants and products are compatible with most real biological systems. 

\subsection{A1c. Influence of the presence of reverse reactions}
\label{app_reverse}

\begin{figure*}
  \includegraphics[width=1\linewidth]{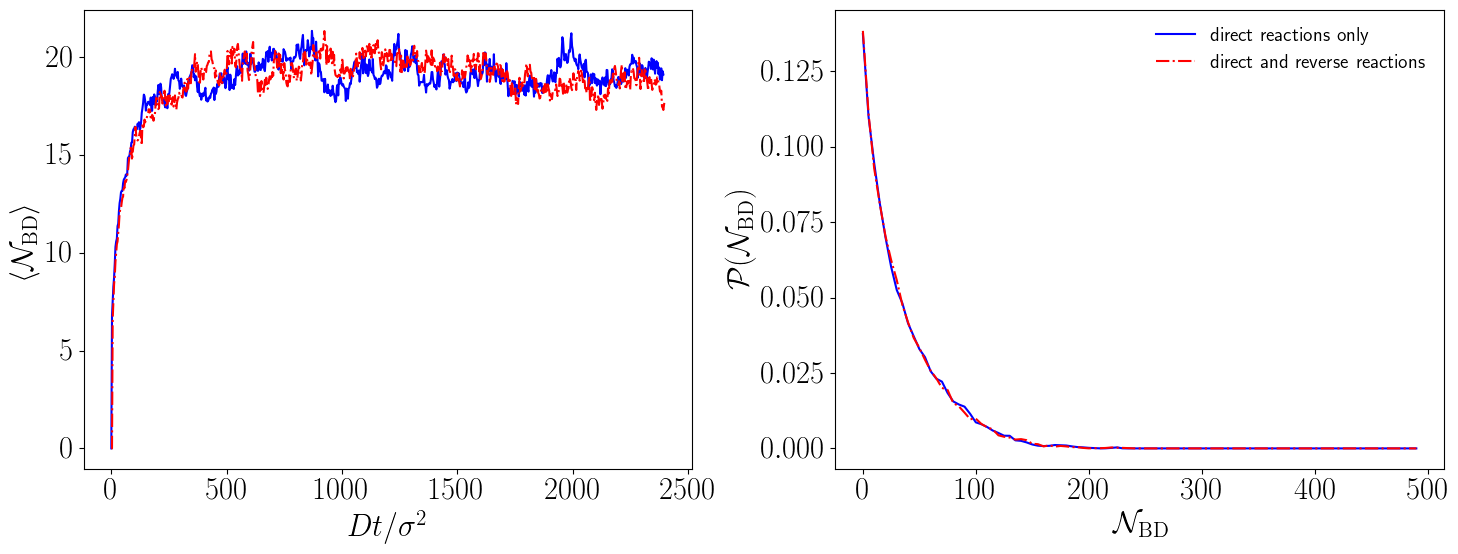}
  \caption{\textbf{Influence of weak reverse reactions in full BD.}  In red, $k_{A\to B}=10$, $k^r_{A\to B}=0.01$ around enzyme $\CAB$, and $k_{B\to A}=10$, $k^r_{B\to A}=0.01$ around enzyme $\CBA$. In blue, $k_{A\to B}=k_{B\to A}=10$, with the reverse reactions neglected, as in the main text. 
  The surface concentration are:  $\rhop=0.1$, $\rhoAB=1.6\ 10^{-3}$, $\rhoBA=3.3\ 10^{-3}$ and $\rhoC=5.3\ 10^{-3}$. Left: Average number of $B$ particles in a droplet as a function of time. Right: Probability density of the number of particles in a droplet at stationary state, computed here from time $5000$ to time $10000$.  }
\label{sup_fig_1}
\end{figure*}

In the study described in the main text, we considered that each type of enzyme only allows a forward reaction in its vicinity ($A \to B$ in the vicinity of  $\CAB$ enzymes, and $B \to A$ in the vicinity of $\CBA$ enzymes). In order to test this assumption, we here consider a series of simulations for which the reverse reaction $B\to A$ with a rate $k^r_{A\to B}$ occurs also in the vicinity of enzyme $\CAB$ (and respectively, the reverse reaction $A\to B$ with a rate $k^r_{B\to A}$ in the vicinity of $\CBA$). 
Nevertheless, we keep considering systems in which the $B$ state is predominant close to $\CAB$ enzymes and 
the $A$ state is predominant close to $\CBA$ enzymes.
Therefore, the rates for the reverse reactions are significantly smaller than the rates of forward reactions ($k^r_{B\to A}=k^r_{A\to B}=0.01$, whereas $k_{A\to B}= k_{B\to A}=10$). As it is shown in Fig.~\ref{sup_fig_1}, the time evolution of the droplet size, and the distribution of the droplet size at stationary state are unaffected by the presence of slow reverse reactions. This justifies neglecting the reverse reactions.

\subsection{A1d. Influence of the range of action of the enzymes}
\begin{figure*}
    \centering
    \includegraphics[width=0.5\linewidth]{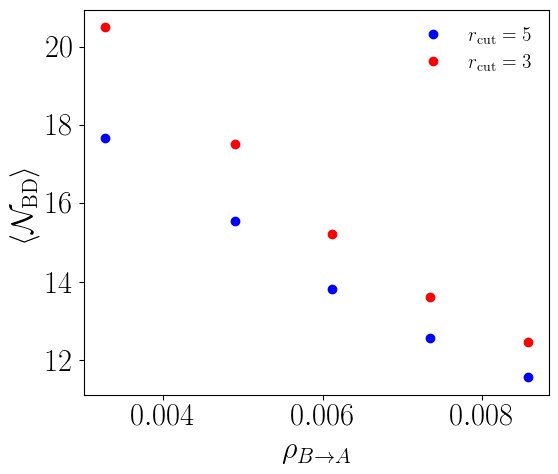}
    \caption{\textbf{Influence of the range of action of enzymes in full BD}.  Average number of $B$ particles in a droplet at stationary state as a function of the surface concentration of $\CBA$ enzymes, for two different values of the parameter $\rcut$, that defines the distance from the center of enzymes under which reactions can occur. Here, $\rhoAB=1.6\ 10^{-3}$. }
    \label{fig:influence_rcut}
\end{figure*}

The range of action of the enzymes is controlled by the parameter $\rcut$. 
Fig.~\ref{fig:influence_rcut} presents the average size of droplets obtained with two different values of $\rcut$, all other parameters being unchanged, as a function of the surface concentration of $\CBA$ enzymes. The results differ quantitatively, but they are qualitatively similar. The larger the value of $\rcut$, the smaller the size of the droplets at the stationary state, as their size is limited by the encounter with $\CBA$ enzymes.

\subsection{A1e. Influence of the size of the simulation box}
\label{app_finite_size}

\begin{figure*}
  \includegraphics[width=1\linewidth]{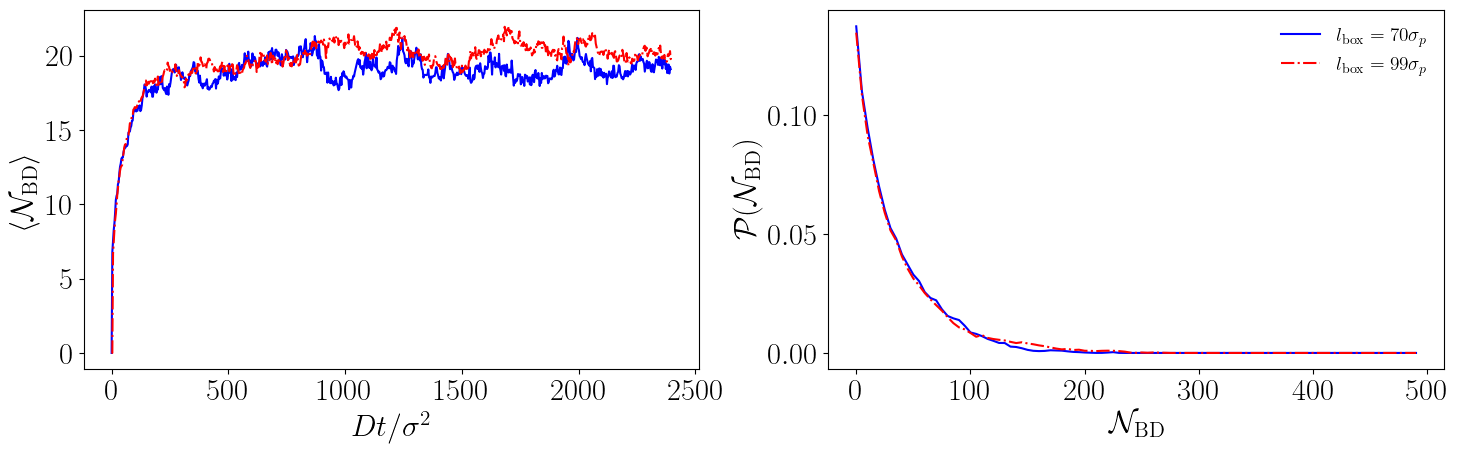}
\caption{\textbf{Influence of the box size in full BD.} 
In red, $\lbox=99 \sigma_p$. In blue, $\lbox=70 \sigma_p$, as in the main text. 
  The surface concentration are:  $\rhop=0.1$, $\rhoAB=1.6\ 10^{-3}$, $\rhoBA=3.3\ 10^{-3}$ and $\rhoC=5.3\ 10^{-3}$. Left: Average number of $B$ particles in a droplet as a function of time. Right: Probability density of the number of particles in a droplet at stationary state, computed here from time $5000$ to time $10000$.}
\label{sup_fig_3}
\end{figure*}

As it is shown in Fig.~ \ref{sup_fig_3}, the time evolution of the droplet size, and the distribution of the droplet size at stationary state, are not strongly affected by the size of the simulation box in full BD. In both cases, surface concentrations of particles are exactly the same, and the results are averaged over the same number of independent realisations. The same trend is found for the time evolution of the droplet size, and for the distribution of droplet size at stationary state, which shows that finite size effects in full BD simularions are negligible. 

\section{A2. Methods: Hybrid dynamics approach (hybrid method, HM)}

\label{appendixHM}
\subsection{A2a. Model}

We consider a binary mixture (BM) of two species $A$ and $B$ characterised by the order parameter $\psi(\rr,t)=\phi_A-\phi_B$, defined as the difference of surface concentration of species $A$ and $B$. It is mixed with a suspension of $N_e$ enzymes, each capable of inducing either the reaction $A \to B$ ($\CAB$ enzyme) or the reaction $B \to A$ ($\CBA$ enzyme) in its vicinity. 
Each enzyme has a finite diameter $\sigma$, and its position is given by $\rr_i$ for $i=1\dots N_e$.

In the absence of reactions, the total free energy of the system, called passive system, is the sum of several contributions, 
\begin{equation}
    F = F_{\rm BM} + F_{ee} + F_{cpl}
\end{equation}
with $F_{\rm BM}$ related to the $A/B$ binary mixture, $F_{ee}$ 
to the enzyme-enzyme interaction,  and $F_{cpl}$ to
the coupling between the field $\psi$ and the enzyme. 
The free energy of the BM is given by a standard Ginzburg-Landau functional of $\psi$  as
\begin{equation}
    F_{\rm BM} = 
    \int d\rr
    \left[
        -\frac{1}{2} \tau \psi^2 +\frac{1}{4} u \psi^4 +
        \frac{1}{2} D (\nabla\psi)^2
    \right]
\end{equation}
where $\tau$ and $u$ specify the local free energy density, while $D$ controls the energetic penalty associated to gradients in the order parameter~\cite{Cahn1959}. 
The minimisation of the local expression of the free energy leads to the equilibrium amplitude $\psi_{eq}$ of the order parameter $\psi$. In bulk regions far from the interfaces, where the gradients of $\psi$ can be neglected, we simply get $\psi_{eq}=\sqrt{\tau/u}$.
Furthermore, both linear stability analysis and minimisation of the free energy can be shown to lead to an equilibrium interface thickness $l_{\rm BM}=\sqrt{D/\tau}$ (related to the largest unstable wavelength).

The enzyme-enzyme free energy contribution is a pairwise additive interaction of a completely repulsive potential that prevents overlapping 
\begin{equation}
    F_{ee} = 
    \sum_{ij} V(r_{ij}/\sigma)
\end{equation}
where $r_{ij}$ is the distance between particles $i$ and $j$.
A soft repulsive Yukawa-like potential is chosen:
\begin{equation}
    V(r)=V_0 \frac{\exp(1-r/\sigma)}{r/\sigma},
\end{equation}
with a cutoff for distances larger than $\sigma$. 

The particle-field interaction writes 
\begin{equation}
    F_{cpl} = 
    \sum_i c \int d\rr \psi_c(|\rr-\rr_i|) \left[ \psi(\rr)-\psi_0 \right]^2
\end{equation}
where $c$ controls the strength of the particle-field interaction and $\psi_0$ controls the selectivity of the particle. 
The tagged function~\cite{Tanaka2000}
\begin{equation}
    \psi_c(r)=\exp\left[ 1-1/(1-(r/R)^2) \right]
\end{equation}
decays smoothly to zero at the corona of the enzyme for $r=R=\sigma/2$.  
The position $\rr_i$ of the particle will minimise the coupling free energy by segregating to regions of space where $\psi(\rr)\approx \psi_0$. 
The strength of the interaction can be quantified by the parameter with energy units $\varepsilon_{cpl}=c\psi_{eq}^2 R^2$.


The dynamics of the binary mixture is controlled by the Cahn-Hilliard-Cook~\cite{Cahn1959,Cahn1959-2,Cook1970} diffusive equation for the order parameter $\psi$ with the addition of a reactive flux arising from the presence of enzymes.
The two coupled dynamic equations for both the field $\psi$ and enzyme position $\rr_i$ are 
\begin{subequations}
    \begin{equation}
        \frac{\partial \psi}{\partial t}=
        M \nabla^2 \left( \frac{\delta F}{\delta \psi} \right)
        +\sqrt{2k_BT~M}\xi
        +\nabla\cdot \jj_a
    \end{equation}
    \begin{equation}
        \frac{d\rr_i}{dt} = 
        \gamma_t^{-1} \ff_i +\sqrt{2D_t} \xi_t(t)
        \label{eq:brownian}
    \end{equation}
    \label{eq:hybrid.model}
\end{subequations}
where $M$ is a mobility constant for the BM~\cite{Barclay2019}, which allows to define a diffusive time scale $t_{\rm BM}=l_{\rm BM}^2/(\tau M)$. 
A random noise $\xi$ is added to the Cahn-Hilliard equation, satisfying fluctuation-dissipation theorem~\cite{Ball1990}. 
A diffusion coefficient $D_t$ is related to the friction $\gamma_t=k_BT /D_t$ and with a noise $\xi_t$ satisfying fluctuation-dissipation theorem. 
Two forces act on the enzyme $\ff_i=\ff_i^{cpl}+\ff_i^{ee}$, 
respectively due to the coupling with the field and the repulsive enzyme-enzyme interaction.

The flux $\jj_a$ is due to the reactions in the vicinity of enzymes. Each enzyme induces a steady conversion rate $K$ in a region of radius $R_a$ centered around $\rr_i$ for enzyme $i$:
\begin{equation}
    \nabla\cdot\jj_a =
    \sum_i K \Theta(R_a-r) 
\end{equation}
where $r$ is the distance between the point in space $\rr$ and the position of the $i$th enzyme $\rr_i$. 
The sign of $K$ depends on the type of reaction that takes place within the $r<R_a$ region: 
if $K>0$, a $\CBA$ enzyme transforms $B$ ($\psi<0$) into $A$ ($\psi>0$); 
if $K<0$, a $\CAB$ enzyme transforms $A$ ($\psi>0$) into $B$ ($\psi<0$). 
The sign of the reaction rate $K$ depends on the local average value of $\psi$ in the vicinity of the enzyme, such that, 
\begin{equation}
    K(\psi_a) = \frac{1}{2} K_0 \left[  1 -{\rm sign}(K_0) \psi_a\right]
\end{equation}
where $\psi_a=\langle \psi \rangle_{r<R_a}$ is the average value of $\psi$ within the $r<R_a$ region for a given enzyme. 
This means that, for an $\CBA$ enzyme producing the $A$ protein, the reaction rate is $K\sim K_0$ in regions where $\psi<0$ ($\psi_a\sim -1$), while $K\sim 0$ in regions where $\psi>0$ ($\psi_a\sim+1$), where no reactant is present.

\subsection{A2b. Dimensionless form of the equations and parameter values}

Due to the large number of parameters in eq. \ref{eq:hybrid.model}, it is useful to express the two coupled dynamic equations in dimensionless form, as 
\begin{subequations}
    \begin{equation}
        \begin{split}
        \frac{\partial \psi}{\partial t}
        = & 
        \nabla^2 
        \left[
        -\psi+\psi^3 -\nabla^2\psi+2\tilde{c}\psi_c(r/\tilde{R})\left(\psi-\tilde{\psi_0} \right)
        \right] \\
        &+\sqrt{2\varepsilon} \xi+ \tilde{t}_K^{-1}\Theta(\tilde{R}_a-r) 
        \end{split}
    \end{equation}
    \begin{equation}
        \frac{\partial \rr_i}{\partial t}=
        \tilde{D_t}\left( 
        \tilde{\varepsilon}_{cpl} \ff_i^{cpl}+
        \tilde{\varepsilon}_{ee} \ff_i^{ee}+
        \right)
        +\sqrt{2\tilde{D}_t}\xi_t(t)
    \end{equation}
\end{subequations}
where length are in units of $l_{\rm BM}$, time in units of $t_{\rm BM}$ and the order parameter amplitude is scaled with $\psi_{eq}$.
The dimensionless parameters verify:
\begin{align}
&\tilde{c}=c/\tau, \;\; \tilde{\varepsilon}_{cpl}=\varepsilon_{cpl}/k_{\rm B}T,  \;\; \tilde{R}=R/l_{\rm BM}, \\
&\tilde{\psi}_0=\psi_0/\psi_{eq}, \;\;
\tilde{t}_K=\frac{t_K}{t_{\rm BM}},  \;\;
\tilde{R}_a=R_a/l_{\rm BM} \\
&\tilde{D}_t = D_t/ M, \;\; \tilde{\varepsilon} = k_{\rm B}T / ( \tau \psi_{eq}^2 l_{\rm BM}^2 )
\end{align}

The simulation box is a squared grid of $256\times 256$ points (except for Fig.~\ref{fig_6_vitesse_enzyme} with a grid of $128 \times 128$ points). The length scale is set such that $l_{\rm BM}=0.85$.
Then, we take for the enzyme radius $\tilde{R}=1.47$, so that the size of the enzyme is comparable to the interface width between $B$-rich droplets.  
The strength of the coupling with respect to enzyme thermal energy, $\varepsilon_{cpl}=240$, and with respect to the BM local energy, $\tilde{c}=2.8$, are chosen to be large enough to ensure that the respective forces and chemical potential are dominant. 
The selectivity of both types of enzyme is $\tilde{\psi}_0=+1$, to mimic the fact that, in equilibrium ($K=0$) all enzymes are dispersed outside of the droplets ($B$ species), \textit{i.e.} in contact with the $A$ phase.  
The diffusivity of the enzyme compared to that of the BM is $\tilde{D}_t=0.2$, to have both components diffusing in a similar time scale. 
The rate of reaction is $\tilde{t}_{K}=0.1$, and the radius of the reaction area is $\tilde{R}_a=9.4$
The scale of the thermal fluctuations $\tilde{\varepsilon}=0.05$ is chosen so that they are subdominant with respect to the characteristic local energy of the BM. 

All simulations are initialised from a random distribution of $\psi$ values centered around $\langle\psi\rangle=0.3$ to ensure the formation of $B$ ($\psi<0$) droplets in a matrix of $A$ ($\psi>0$). 
A droplet is defined as the system region $(x,y)$ where $\psi(x,y)<0$. Standard cluster analysis are used to determine the individual droplets, to then characterise the number of droplets and the mean droplet size, defined as the average number of grid points in droplets, rescaled by the unit area $l_{\rm BM}^2$. A droplet is defined as a cluster of at least $5$ grid points.

\section{A3. Analytic model for droplet growth}
\label{annex_analytical_model}

We propose a simple model to describe the evolution with time of the mean droplet size, which largely differs from the prediction of the Oswald ripening mechanism. We consider $n$ independent droplets, each made of $\mathcal{N}$  proteins, and connected to a reservoir of $B$ proteins.

The reservoir yields a constant and equal influx $c$ of proteins towards each droplet. The droplets are assumed to be totally emptied when they encounter an $\CBA$ enzyme, and restart growing just after the encounter. Consequently, the evolution equation for each droplet, between $t$ and $t+\dd t$ is:
\begin{align}
    \mathcal{N}(t+\dd t)=\left\{
    \begin{array}{ll}
        c\ \dd t & \mbox{if encounter in [$t$,$t + \dd t$]}\\
        \mathcal{N}(t)+c\ \dd t & \mbox{otherwise.}
    \end{array}
\right.
\label{eq_evolution_N_i}
\end{align}
We assume now that the encounters between $\CBA$ enzymes and droplets are memoryless, so that encounter times follow an exponential probability distribution. We call $\tau$ the expectation value of this distribution, corresponding to the average encounter time between a droplet and an $\CBA$ enzyme. 
Therefore, the probability that the encounter takes place between $t$ and $t+\dd t$ is $\dd t/ \tau$.

By averaging Eq.~\ref{eq_evolution_N_i} , we get:
\begin{align}
    \langle \mathcal{N}(t+\dd t) \rangle 
    &=(\langle \mathcal{N}(t) \rangle +c\ \dd t)(1-\frac{dt}{\tau}) +c\ \dd t \frac{\dd t}{\tau}\\
    &= \langle \mathcal{N}(t) \rangle(1-\frac{dt}{\tau}) +c\ \dd t
\end{align}
Thus,
\begin{align}
    \frac{\partial \langle \mathcal{N} \rangle}{\partial t} 
    &=-\frac{\langle \mathcal{N} \rangle}{\tau} +c
\end{align}

Before solving this equation, we express $\tau$ in terms of the parameters of the model. 
If we neglect correlations between $\CBA$ enzymes (low concentration regime), this characteristic encounter time can be approximated by the characteristic encounter time $t_e$ between a droplet and a single $\CBA$ enzyme, divided by the number of $\CBA$ enzymes $\mathcal{S}\rho_{B\to A}$, i.e. $\tau={t_e}/\mathcal{S}\rho_{B\to A}$ (with $\mathcal{S}$ being the area of the simulation box and $\rho_{B\to A}$ being the surface concentration of $\CBA$ enzymes). Also,  $t_e$ depends on the diffusion coefficients of both particles, and thus verifies:
\begin{align}
    t_e \propto \frac{1}{D_{\rm dp}+D_{\CBA}},
\end{align}
with $D_{\CBA}$ the diffusion coefficient of the $\CBA$ enzyme, and $D_{\rm dp}$ the diffusion coefficient of the droplet.

Solving this differential equation with initial conditions $\langle \mathcal{N} \rangle=0$, and using the previous expression of $\tau$, yields:
\begin{align}
    \langle \mathcal{N} \rangle=c\frac{t_e}{\mathcal{S}\rho_{B\to A}} \left(1-\exp\left(\frac{-t \mathcal{S} \rho_{B\to A}}{t_e}\right)\right)
\end{align}
This corresponds to Eq. \ref{expression_analytique_N} in the main text.

\enlargethispage{20pt}

%

\end{document}